\documentclass[10pt]{article}
\usepackage[margin=1in]{geometry}

\usepackage{setspace} 
\usepackage{lineno} 

\usepackage{graphicx}
\usepackage{amssymb}
\usepackage{amsmath}
\usepackage{epstopdf}

\usepackage{natbib}

\usepackage{sidecap} 
\usepackage{color}
\usepackage{amsmath}
\usepackage{mathabx}
\usepackage{bbold}

\newcommand{\rd}{{\rm d}}
\newcommand{\E}{{\rm E}}

\newcommand{\ns}{{n_{\rm s}}}
\newcommand{\nx}{{n_{\rm x}}}
\newcommand{\nG}{{n_{\rm g}}}

\newcommand{\pD}{{p_D}}

\renewcommand{\vec}[1]{{\ensuremath{ \boldsymbol{ \bf #1}}}}
\newcommand{\vx}{\vec{x}}
\newcommand{\vI}{\vec{I}}
\newcommand{\vy}{\vec{y}}

\newcommand{\vg}{\vec{g}}
\newcommand{\Propto}{\;\propto\;}

\newcommand{\phig}{\phi^{(\vg)}}

\newcommand{\appropto}{\mathrel{\vcenter{
  \offinterlineskip\halign{\hfil$##$\cr
    \propto\cr\noalign{\kern2pt}\sim\cr\noalign{\kern-2pt}}}}}

\title{The implications of perception as probabilistic inference for correlated neural variability during behavior}

\author{
Ralf M.~Haefner$^1$
\thanks{ralf.haefner@gmail.com}, \;   
Pietro Berkes$^2$ and J\'ozsef Fiser$^3$\\
$^1$ \small Brain \& Cognitive Sciences, University of Rochester, Rochester, NY 14627, USA\\
$^2$ \small Sloan-Swartz-Center for Theoretical Neurobiology, Brandeis University, Waltham MA 02454, USA\\
$^3$ \small Department of Cognitive Science, Central European University, Budapest 1051, Hungary}

\begin{document}

\maketitle

\begin{abstract}
This paper addresses two main challenges facing systems neuroscience today: understanding the nature and function of a) cortical feedback between sensory areas and b) correlated variability. Starting from the old idea of perception as probabilistic inference, we show how to use knowledge of the psychophysical task to make easily testable predictions for the impact that feedback signals have on early sensory representations. Applying our framework to the well-studied two-alternative forced choice task paradigm, we can explain multiple empirical findings that have been hard to account for by the traditional feedforward model of sensory processing, including the task-dependence of neural response correlations, and the diverging time courses of choice probabilities and psychophysical kernels. 
Our model makes a number of new predictions and, importantly, characterizes a component of correlated variability that represents task-related information rather than performance-degrading noise. It also demonstrates a normative way to integrate sensory and cognitive components into physiologically testable mathematical models of perceptual decision-making.
\end{abstract}

\section*{Introduction}

Almost 150 years ago, Helmholtz proposed that visual perception is an inference process  \cite[]{Helmholtz1866}. He suggested that at any point in time the brain combines its prior knowledge about the external world with the incoming sensory information to compute the most likely explanation for its inputs. Substantial evidence from both psychophysics  \cite[]{Kersten2004} and recent physiological findings \cite[]{Berkes2011} suggests that the responses of visual neurons are influenced by such prior information, also called an 'internal model'. 
Since sensory information is usually uncertain and often incomplete, this proposed inference and the assumed internal model both need to be probabilistic in nature  \cite[]{Fiser2010, Pouget2013}. 
Furthermore, it has been noted that probabilistic inference in a hierarchical model requires a flow of information remarkably similar to the one within the visual system: feedforward from the retina; and feedforward, recurrent and feedback within cortex  \cite[]{Mumford1992,Lee2003}. 
However, the major challenge to generating neurophysiologically testable predictions based on probabilistic inference, and for incorporating feedback in this framework has been that the nature of the internal model that the brain uses for general vision is currently unknown.

\begin{figure}\begin{center}
  \includegraphics[width=10cm]{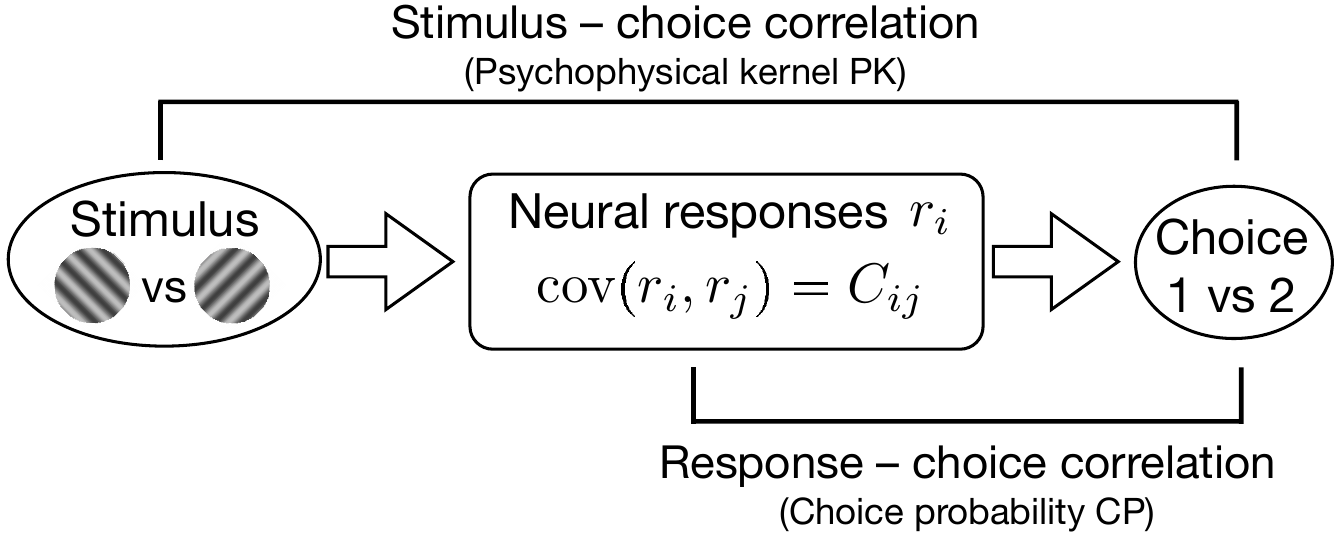}
\caption{Schematic of the perceptual decision-making system modeled in this paper. Information about  a visual stimulus is represented by the responses $r_i$ of a population of sensory neurons whose activity influences the behavioral choice of the subject. This paper focuses on the following three observables in this system: 
(i) The structure of the correlations between neural responses (noise correlation) believed to be important for the amount of information that can be represented by neural responses  \cite[]{Zohary1994,Abbott1999,Ecker2011}.
(ii) The time-course of the magnitude of the choice-triggered stimulus average, also called psychophysical kernel (PK), which quantifies the influence of the stimulus on the final decision as a function of the time when the stimulus is presented during the trial  \cite[]{Neri1999, Ahumada2002, Nienborg2009, Brunton2013}.
(iii) The correlation between a single neuron's response and the final choice, usually quantified as choice probability (CP)  \cite[]{Britten1996,Nienborg2012,Haefner2013}.
}
\label{fig_Setup}
\end{center}\end{figure}

We overcame this challenge by studying the consequences of probabilistic inference in a well-controlled task in which the generative model of the sensory inputs is under the control of the experimenter. For the subject to perform probabilistic inference in such a task, its brain needs to learn this experimenter-defined generative model. Learning such a model can be interpreted as a perturbation on the unknown internal model for general vision. We can therefore use our knowledge of this experimenter-defined perturbation to make predictions about how the neural responses should change as a result of it. We applied these ideas to the well-studied two-alternative forced choice  (2AFC) task paradigm and, assuming that the brain performs inference by 'neural sampling'  \cite[]{Hoyer2003,Fiser2010}, we generated predictions for both the responses of sensory neurons and for psychophysical measurements.
We find that our probabilistic inference model correctly reproduces key experimental observations in 2AFC tasks: the task-dependency of noise correlations  \cite[]{Cohen2008}, the temporal increase of the correlation between sensory responses and behavior (called choice probability, CP  \cite[]{Cohen2009, Nienborg2009, Nienborg2012}) and the decrease in the correlation between stimulus and behavior (called psychophysical kernel, PK \cite[]{Neri1999, Ahumada2002}) in some tasks  \cite[]{Nienborg2009}, but not others  \cite[]{Brunton2013} (Figure \ref{fig_Setup}). 

We also derived a number of specific predictions based on our model that can readily be tested in empirical studies. First, we predict that the noise correlations for sensory responses in a 2AFC task have two maxima and two minima whose locations are defined by the task-relevant stimuli. We further predict an increase in the amplitude as defined by those maxima and minima on the time scale of perceptual learning. Psychophysically, we predict that the relative weighting of evidence throughout the trial should depend on stimulus strength. Since our model reflects the structure of the task, and was not designed to fit any of the existing observations made in specific experiments, we predict that our findings generalize to other stimuli, modalities and sensory areas in the brain.  
In general, the fact that our framework's predictions directly reflect the spatial and temporal structure of the experimenter-defined task makes them easy to test. Importantly, since its predictions concern the full statistical structure of neural responses, including higher level correlations, our framework provides theory-driven guidance on how to analyze the high-dimensional data obtained from increasingly common population recording techniques.

\section*{Results}

A basic assumption in our framework is that sensory neurons in cortex represent the brain's belief about some aspect of the outside world \cite[]{Lee2003}. In the case of V1, we assumed this to be how well a Gabor-shaped feature describes the image on the retina at the location of the neuron's receptive field \cite[]{Olshausen1997}. From that follows that the activity of V1 neurons will depend \emph{both} on the image on the retina \emph{and} on any prior knowledge about the outside world in the rest of the brain. The former, called likelihood in probabilistic inference, is communicated to V1 via feedforward connections, while the latter, the prior, arrives via feedback connections. Training a subject on a given psychophysical task will induce such prior knowledge in the subject's brain about the stimulus that is being presented. In this paper, we will show how to use knowledge about the task structure to make predictions about its influence on V1 responses. 

\subsection*{Qualitative intuitions}

Let's assume a simple coarse orientation discrimination task where the subject has to report whether \emph{either} a vertical \emph{or} a horizontal grating is embedded in a noisy image. Then on any given trial, if the subject believes the vertical grating to be present, neurons representing vertical orientations will be more active than baseline, and neurons representing horizontal orientations will be less active -- \emph{even if there is no signal in the stimulus at all}. This means that those neurons will be correlated with the choice (commonly quantified as choice probabilities, Figure \ref{fig_CP}b). Furthermore, as a side-effect of this correlation between neural response and perceptual decision, the correlations between the responses of neurons supporting the same choice will be higher than between the responses of neurons supporting opposing choices (Figure \ref{fig_Corr}). Finally, since the internal belief about the correct choice develops gradually by integrating the sensory information over the course of a trial, the magnitude of those correlations should also increase over the course of a trial (Figure \ref{fig_CP}). As a side-effect, the dynamic feedback about the current belief about the correct choice on the sensory responses may -- in the case of very weak external inputs -- lead to a self-reinforcing loop with the result that early evidence has a bigger influence on the final decision than late evidence (Figure \ref{fig_PK}).

\subsection*{The model}

In order to make the above intuitions quantitative, we will proceed in three steps. First, we need to make an assumption about the generative model that the brain has learnt and performs inference in. Second, we need to make an assumption about how the variables in this model are related to sensory neurons. And third, we need to specify how the brain's beliefs are related to the responses of those neurons.

\begin{figure}\begin{center}
  \includegraphics[width=14cm]{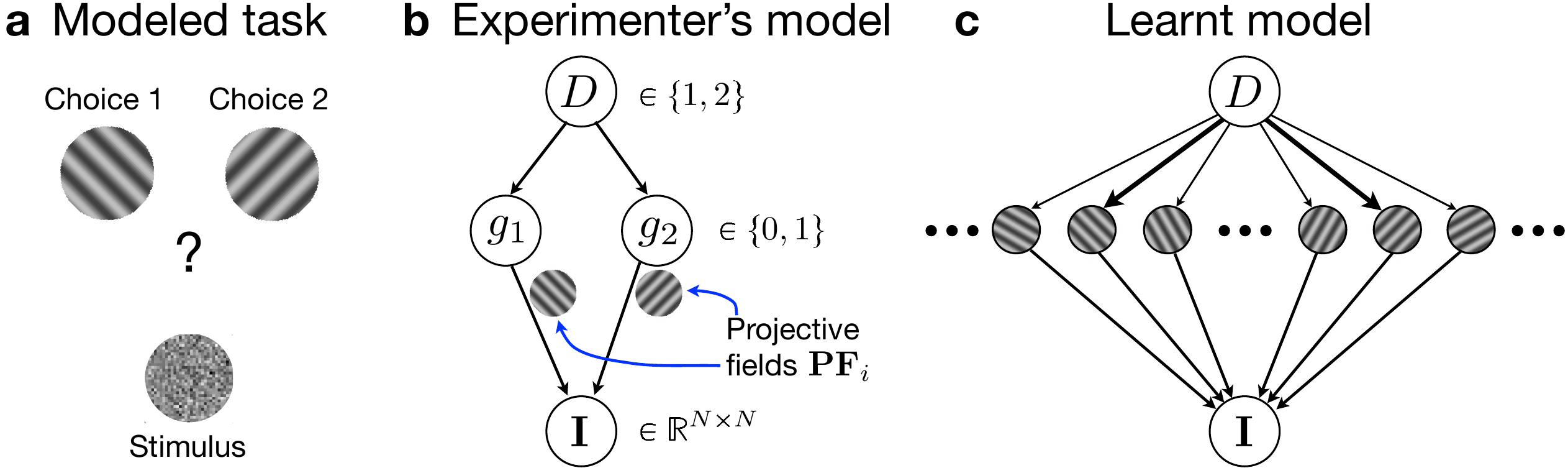}
\caption{The structure of the task.
{\bf a:} The task is to decide whether a noisy display was caused by a $-45$deg or $+45$deg oriented grating.
{\bf b:} Experimenter's model. First the rewarded decision $D$ is chosen from some prior distribution, $p(D)$. Depending on the value of $D$, $1$ or $2$, the corresponding grating variable $g_1$ or $g_2$ is set to $1$ while the value of the other grating variable is set to zero. The image $\vI$, consisting of $N\times N$ pixels on the screen, is drawn from some noisy distribution around the signal/target images, $g_1\vec{T}_1+g_2\vec{T}_2$.
{\bf c:} Model learnt by the brain. Due to the noise in the stimulus, the correct decisions corresponding to the two values of $D$ may be caused not just by the gratings with exactly $-45$deg or exactly $+45$deg orientation but, with smaller probability, also by gratings of similar orientation (indicated by arrows of different width.) 
The relationships between $D$ and the $g_i$'s are not deterministic but probabilistic and learnt by the subject by correlating the rewarded responses with the stimuli preceding them. Note that arrows here and in all following figures other than Figure \ref{fig_Combined} indicate statistical dependencies in a Bayesian network, {\it not} information flow (which is shown in Figures \ref{fig_Combined}b and \ref{fig_PK}c).
}
\label{fig_GenModel}
\end{center}\end{figure}

The generative model of a psychophysical task consists the relationship between the available behavioral choices and the stimulus, as well as the prior probabilities of each choice being the correct one.
In 2AFC tasks the behaviorally relevant variable is binary and related to one of two possible choices. In visual tasks, the sensory input is the luminance pattern on the retina. For concreteness, we present our results in the context of an orientation discrimination task (Figure \ref{fig_GenModel}a). However, our results are also applicable to 2AFC tasks based on other stimuli (e.g. motion, disparity) or modalities (e.g. auditory, vestibular). We compared our model predictions with empirical data from a coarse motion direction discrimination task  \cite[]{Cohen2008,Cohen2009}, a coarse disparity discrimination task  \cite[]{Nienborg2009} and an auditory task  \cite[]{Brunton2013}.

First, the correct decision, $D$, is drawn from a prior distribution (here we assumed $p(D=1)=p(D=2)=0.5$) (Figure \ref{fig_GenModel}b). $D$ determines which one of two target images (here grating images) will be embedded in Gaussian noise. We represent the presence or absence of each target by $g_1$ and $g_{2}$ being $0$ or $1$, respectively. 
Probabilistic inference in this task entails computing the probability distribution over the relevant unobserved variables (here: $D$), given the observed ones (here the image $\vI$): $p(D|\vI)$. While the experimenters control the image $\vI$, they learn about the subject's belief about $D$ in any one trial by recording the subject's choice.

The subject achieves optimal performance if the internal model that its brain has learnt agrees with the actual (external) model that generates the stimulus defined by the experimenter.
However, while the experimenter's model defines the theoretical optimum, the model learnt by the brain is likely to deviate from it.\cite[]{Ma2012}  Here, we model two such deviations. The first deviation is due to the fact that the brain has to learn the task by correlating rewarded choices with preceding noisy stimuli.\cite[]{Qamar2013} As a result, there will be some uncertainty about the exact task-relevant orientations (Figure \ref{fig_GenModel}c). Furthermore, while the relationships between $D$ and $g_{i}$ are deterministic in the experimenter's model such that $p(g_{1}=1|D=1)=1$ and $p(g_{2}|D=1)=0$, for instance, allowing for other explanations for the image on the retina will change that probability to intermediate values between $0$ and $1$. Our assumptions about these deviations manifest themselves as two free parameters in our model (Methods). 

\begin{figure}\begin{center}
  \includegraphics[width=8.5cm]{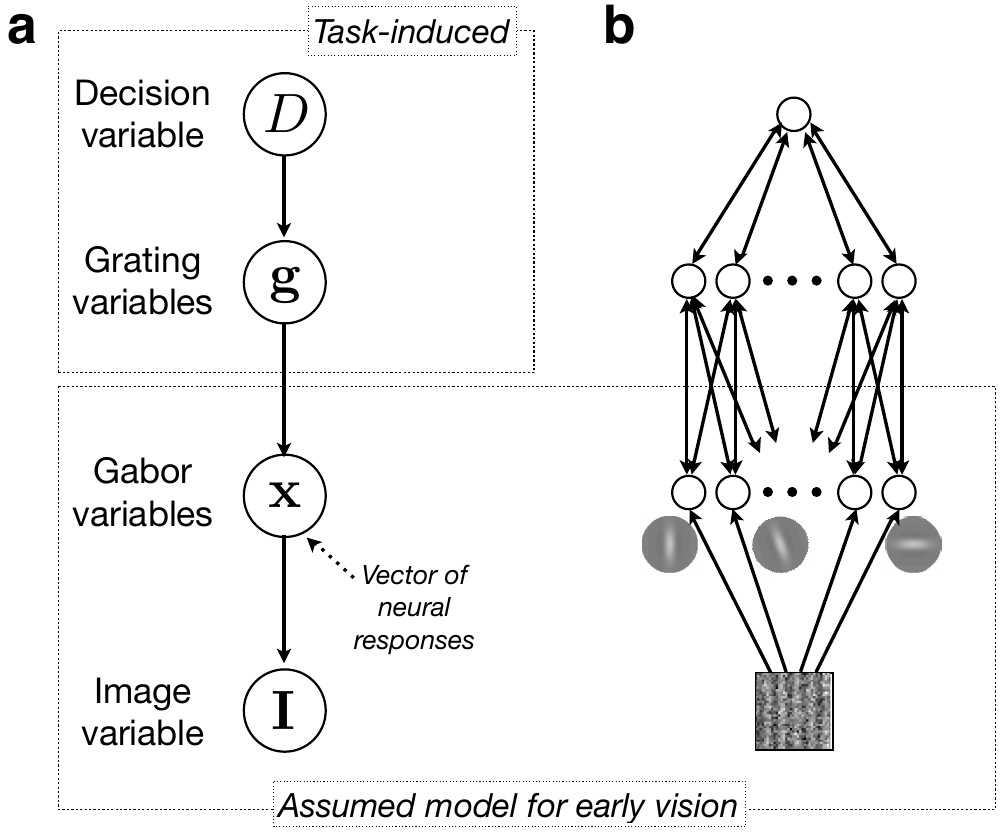}
\caption{Full probabilistic model consisting of a sparse coding model for early vision  \cite[]{Olshausen1996, Hoyer2003} and the generative model for the task.
{\bf a:} Graphical representation of the combined generative model.
{\bf b:} Correspondence to biology. We assume $\vx$ to be represented by early visual neurons, for instance in V1. While our results do not depend on the details of the implementation of grating and decision variables in the brain, we imagine them being represented in mid or high-level visual areas and decision-making areas, respectively.
}
\label{fig_Combined}
\end{center}\end{figure}

In order to make experimentally testable predictions for the neuronal responses in primary visual cortex (V1) we combine the generative model of the task with the probabilistic version of an established sparse-coding model for V1  \cite[]{Olshausen1997, Hoyer2003}. In this model the responses of V1 simple cells, $\{x_i\}$, are assumed to represent the intensity of an oriented Gabor-shaped feature in the image at a particular location (Figure \ref{fig_Combined}, Methods). The idea underlying our approach is a) that the brain will perform the task based on the responses of its existing V1 neurons, and b) that the relationship between retina and V1 is not substantially altered by learning a task during adulthood, long after the critical period. As a result, each variable $g_i$ in our task model now represents the task-relevant responses of neurons in our V1 model, $\{x_i\}$ rather than a pixel pattern on the screen directly (Figure \ref{fig_Combined}, Methods). 
If the brain indeed solves the task by performing probabilistic inference over all unobserved variables given the observed one ($\vI$), the task-structure (which defines the upper part of our model) acts as a prior on the activity of the V1 neurons represented by $\vx$. Hence, task-related knowledge, i.e. expectations about which stimuli are more or less likely in the sensory inputs, is thereby incorporated in the sensory neurons' responses. This top-down influence induces correlations between the V1 neurons (noise correlations), between the neurons and the decision (choice probabilities), and between the stimulus and the decision (psychophysical kernel). Since these three types of correlations are directly observable, we use them to link our model predictions to empirical data.

Finally, we need to specify the algorithm by which the brain performs inference in the generative model defined by Figure \ref{fig_Combined}, and how its beliefs about the variables in this model are represented by neural responses. While both parametric  \cite[]{Ma2006} and sampling-based implementations have been proposed, we assumed a neural-sampling based representation supported by many studies in cognition (both for inference, reviewed in \cite[]{Vul2014}, and learning \cite[]{Bonawitz2014}) and increasingly also sensory processing \cite[]{Hoyer2003,Fiser2010, Berkes2011,Buesing2011,Pecevski2011} (also see \cite[]{Moreno-Bote2011} for an approach combining both parametric and sampling-based representations). In short, the neural sampling hypothesis proposes that the brain performs inference, i.e. computes $p(D,\vec{g},\vec{x}|\vI)$, by generating a sequence of 'samples' from this probability distribution based on the generative model that it has learnt. In our case, a single sample, indexed by $k$, from this probability distribution $p(D,\vg,\vx|\vI)$ is a vector $(D^{(k)},g_1^{(k)},g_2^{(k)},..,x_1^{(k)},x_2^{(k)},..)$. 
In a sampling-based representation, the marginal probability $p(D|\vI)$ can be deduced from the sequence of samples $D^{(k)}$. This means that a decision-making area computing the probability that the display on the screen was caused by one orientation rather than the other one, can simply count the number of occurrences of $D^{(k)}=1$ and $D^{(k)}=2$ over time, for instance by increasing or decreasing the activity of a pool of neurons as has been proposed previously  \cite[]{Gold2007}.

The generative model (Methods) specifies the sampling equations (Supplementary Information) that constitutes a mechanistic model for how each V1 neuron in our model updates its firing rate (represented by $x_{i}$) depending on its inputs from the retina (feedforward), the other V1 neurons (lateral), and the rest of the brain (feedback, specified by the task model). In our simulations, we modeled a trial of 1 second duration by generating 80 samples from the full model as described in Figure \ref{fig_Combined} (see Methods). The 80 samples for a single simulated V1 neuron represent both the neuron's firing rate changing throughout the trial and the brain's evolving belief about the intensity of the corresponding Gabor-shaped feature in the image. The sum of all 80 samples for a particular $x_i$ represents the spike count over the duration of the entire trial used to compute noise correlations for Figures \ref{fig_Corr} and \ref{fig_Corr_2}. Individual samples are used to compute the instantaneous correlation between a neuron's response and the decision for Figure \ref{fig_CP}c.

\subsection*{Task-induced correlations between sensory neurons}

\begin{figure}\begin{center}
  \includegraphics[width=12cm]{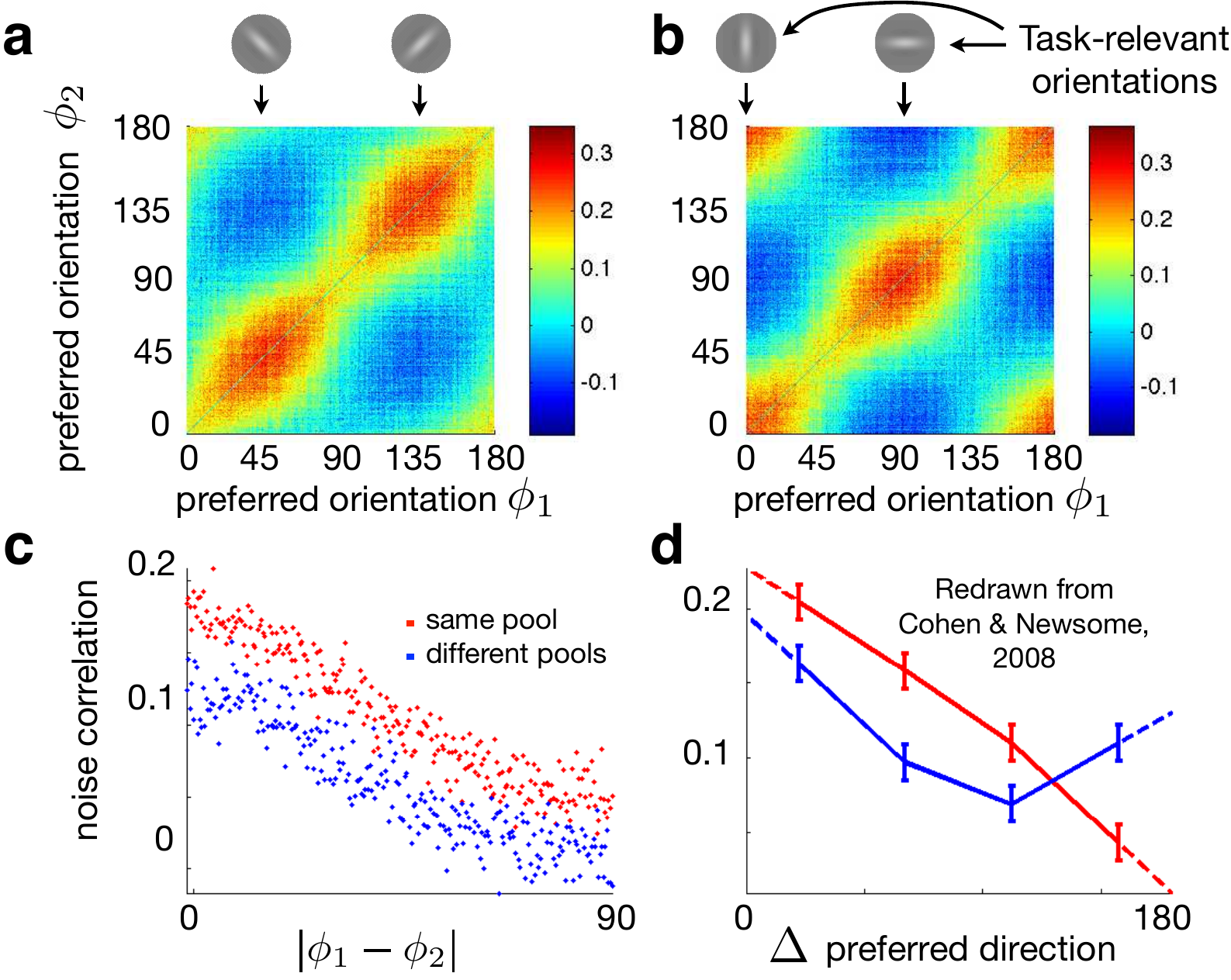}
\caption{Model predictions for noise correlations between the sensory neurons in our model.
{\bf a:} Full correlation matrix with neurons sorted by their preferred orientation. The task-relevant orientations are indicated at $45$deg and $135$deg. Only the qualitative shape is predicted by our model -- the overall magnitude depends on the particular choice of model parameters (and can be fit to data).
{\bf b:} Same as panel {\bf a}, but for a cardinal orientations task.
{\bf c:} Correlation coefficients as a function of difference in preferred orientation between the two neurons in a pair. Red: all pairs where each neuron's preferred orientation is closest to the same task-relevant orientation (i.e. 'supports' the same choice). Blue: all pairs in which the two neurons are aligned with different choices. The variability around the means reflects the 'measurement noise' from 1000 simulated trials.
{\bf d:} Noise correlations recorded between neurons in area MT for a motion direction task  \cite[]{Cohen2008}. Data replotted from Figure 4 therein. Since motion direction has a period of 360deg compared to a period of 180deg for orientation, differences in preferred orientation between 0 and 90deg in panel {\bf c} are comparable to differences in preferred direction between 0 and 180deg in panel {\bf d}.
}
\label{fig_Corr}
\end{center}\end{figure}

\begin{figure}\begin{center}
  \includegraphics[width=12cm]{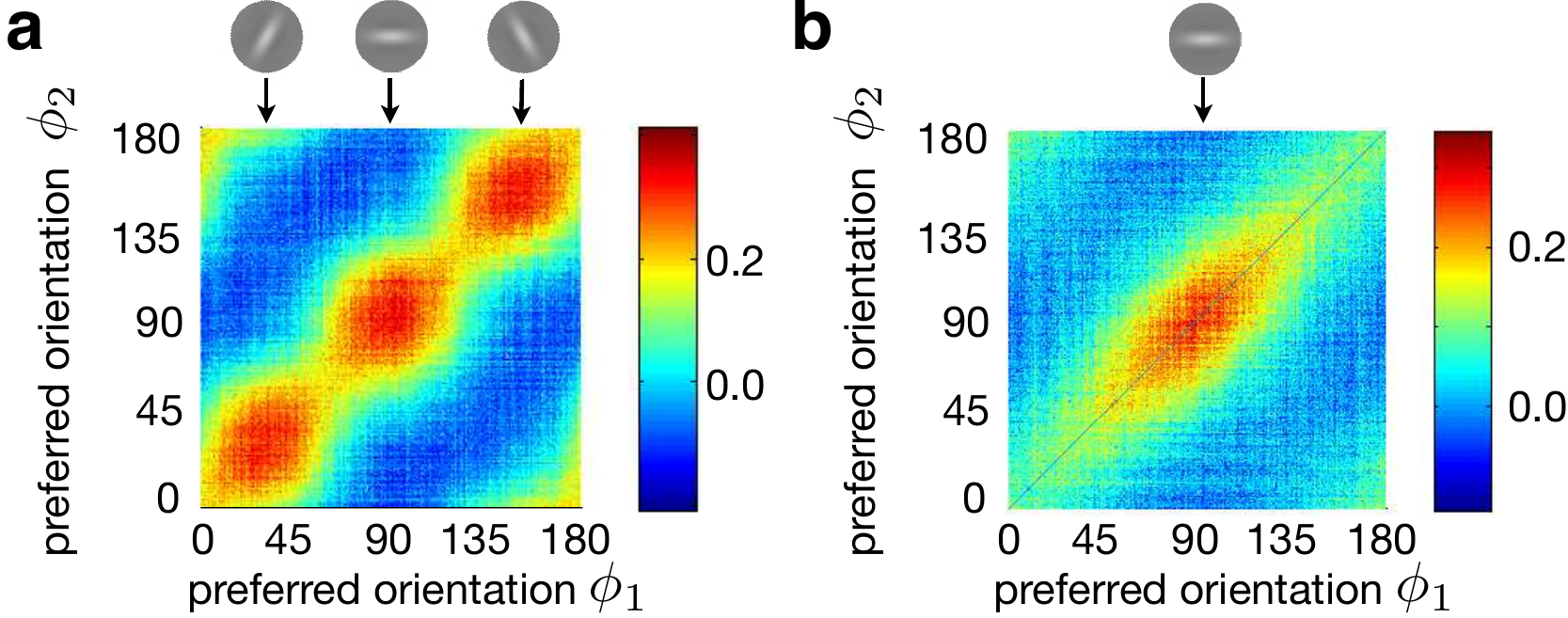}
\caption{Predicted noise correlations for alternative tasks analogous to Figure \ref{fig_Corr}. 
{\bf a:} Three-alternative choice task (3AFC). 
{\bf b:} Detection task.
}
\label{fig_Corr_2}
\end{center}\end{figure}

Figure \ref{fig_Corr} shows the trial-to-trial correlation coefficients between the sensory neurons in our model while the system is performing inference on a stimulus without signal (no grating present) -- commonly called noise correlations  \cite[]{Cohen2011rev, Nienborg2012}. The most important feature to note are the two maxima and the two minima in the correlation structure occurring at the task-relevant orientations: $45$ degrees vs $-135$ degrees (panel a) and vertical vs horizontal (panel b). The correlation peaks are a direct result of the fact that neurons with stimulus preferences close to the same task-relevant stimulus will be increased and decreased \emph{together} as the subject's belief about the correct choice varies from trial to trial. At the same time, neurons who support different choices, will have reduced correlations since their responses are increased and decreased on different trials. This deviation from rotational symmetry is task-dependent in that the locations of peaks and troughs are entirely determined by the task-relevant orientations. Furthermore, our framework can easily be used to predict the noise correlation structure for other psychophysical tasks. For example, for a 3AFC task our framework predicts the noise correlation structure to have three peaks along the diagonal and corresponding three troughs on each side of the diagonal (Figure \ref{fig_Corr_2}a). Similarly, a detection task is predicted to induce a single peak at the task-relevant stimulus (Figure \ref{fig_Corr_2}b).

A signature of this task-dependence of the correlation structure has previously been observed in area MT in a coarse motion direction task  \cite[]{Cohen2008}. In order to facilitate comparison with the data from that experiment, we collapsed our correlation matrix and plotted the correlation coefficient as a function of the difference in the preferred orientations of the two neurons (Figure \ref{fig_Corr}c.) Following Cohen \& Newsome \cite[]{Cohen2008}, we plotted this dependency separately for neuron pairs where both neurons prefer the same task-relevant orientation (red) and for neuron pairs supporting opposing choices (blue). As in the empirical data (Figure \ref{fig_Corr}d), we found a reduction in correlations between neurons supporting different choices compared to pairs of neurons supporting the same choice. Our model prediction deviates from this dataset at the largest differences in orientation. However, it fully agrees with more recent data (Bondy \& Cumming, SfN 2013 Conference Abstract) which does not find a crossover in the correlation profiles at large differences between preferred orientation. Future studies will be needed to understand this difference between datasets since both 2AFC tasks -- one in the motion direction domain \cite[]{Cohen2008} and one in the orientation domain (Cumming \& Bondy, SfN 2013) -- are identical from the perspective of our framework, and hence model. 

The second feature to note in the correlation structure (Figure \ref{fig_Corr}a+b) is the elongated shape of the peaks and troughs. This is due to the task-uncertainty included in our model. The more precisely the task-relevant orientations are known by the subject, the closer to rotationally symmetric they are predicted to be. This prediction is consistent with the observation that a substantial uncertainty is required in our model (parameter $\kappa$, Methods) to find a reasonable agreement with the data of Cohen \& Newsome (2009) where the monkeys are cued to the task-relevant directions on a trial-by-trial basis, while there is very little sign of any elongated shape in their correlation matrix, and hence task-uncertainty, in the data of Bondy \& Cumming (SfN 2013 Conference Abstract) who train their monkeys for several days on every new pair of task-relevant orientations before measuring the correlation matrix.

Our model does not predict the absolute magnitude of the correlation matrix, and hence the overall level of correlations. Any non-task-specific changes in the overall level of excitability (e.g. due to changes in alertness, \cite[]{Ecker2014a}) would add a positive offset to the correlations in Figures \ref{fig_Corr}+\ref{fig_Corr_2}. However, since the amplitude of the correlation matrix (measured as peak minus trough) directly reflects the degree of task-knowledge, another new prediction of our framework is an {\it increase} in the amplitude with learning (Figure \ref{fig_Corr_Performance}). As the brain learns a better approximation of the experimenter's task its prior becomes stronger and the top-down influence increases in our model, a process that may be related to perceptual learning  \cite[]{Goldstone1998}.

\subsection*{Correlations between sensory responses and behavior}

\begin{figure}\begin{center}
  \includegraphics[width=14cm]{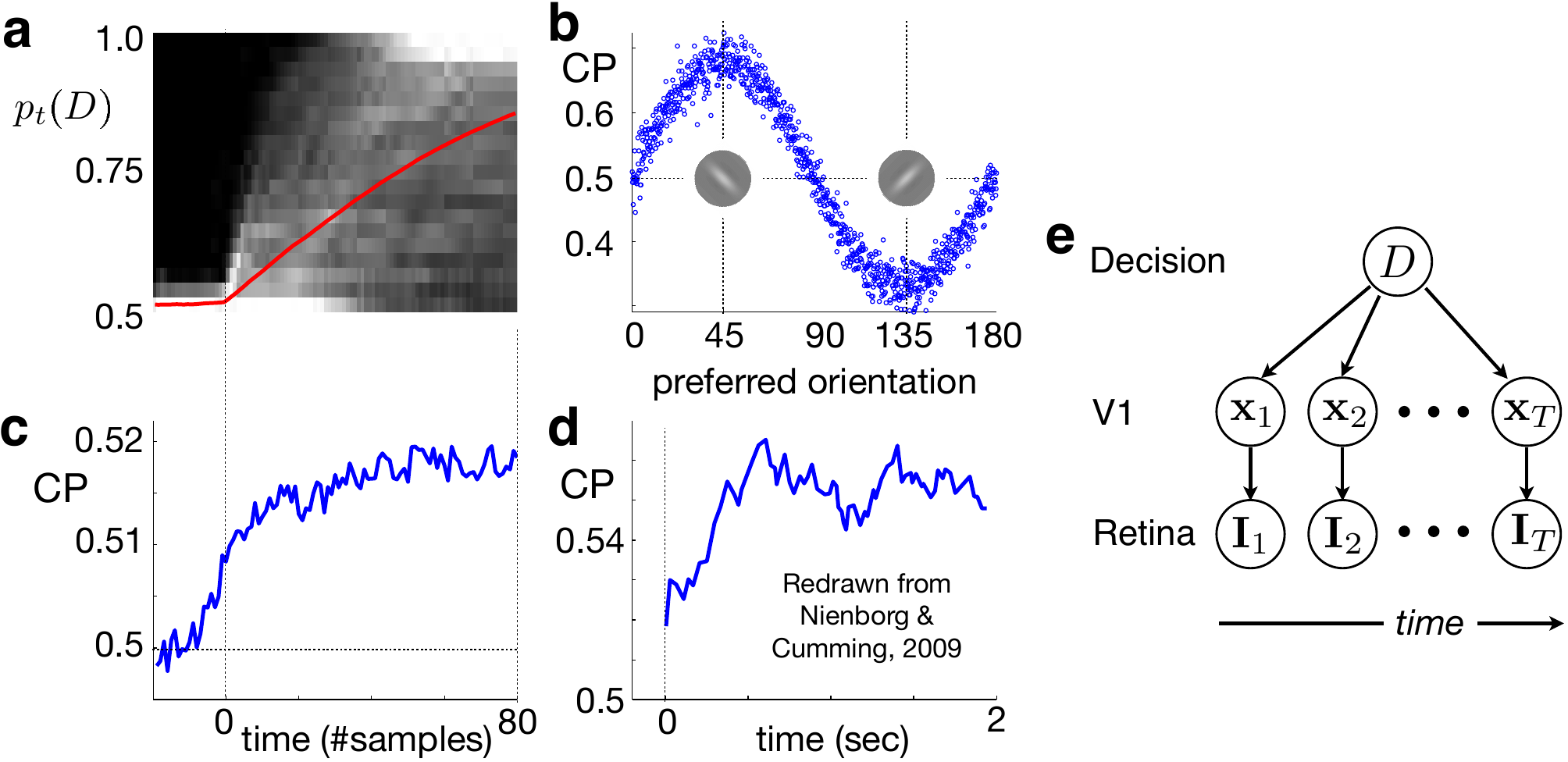}
\caption{Model prediction for choice probabilities (CP). 
{\bf a:} Temporal evolution of the belief about the correct decision $D$. Red curve shows the average posterior $p_t(D=1)$ as a function of time for those trials for which the final choice was $D=1$. It is fixed to $0.5$ until trial onset and accumulates evidence throughout the trial. The grayscale map shows the distribution of beliefs as a function of time (high probability in white, zero probability in black). Each column represents a histogram of $p_t(D=1)$ at time $t$.
{\bf b:} CP as a function of preferred orientation. Here, CPs were computed across the entire trial rather than in small time bins as in {\bf c} which is why their magnitude is greater than in panel c. 
{\bf c:} Magnitude of CP as a function of time. 
{\bf d:} CP magnitude as a function of time based on data recorded in area V2 for a disparity direction task. \cite[]{Nienborg2009}.
{\bf e:} Generative model used for the task extended into the time-domain (Figure \ref{fig_Combined}a after integrating out the grating variables). 
}
\label{fig_CP}
\end{center}\end{figure}

Another much-studied empirical quantity in the context of perceptual decision-making is 'choice probability' (CP)  \cite[]{Britten1996,Shadlen1996,Haefner2013,Nienborg2012}, essentially a measure for the correlation between the variability in the response of a sensory neuron and the behavioral choice in each trial (Methods). Our model makes three predictions in agreement with existing empirical data. First, it predicts CPs to grow as a function of time (Figure \ref{fig_CP}c, compare with data redrawn in Figure \ref{fig_CP}d). Whether CPs reflect the causal influence of a sensory neuron on the decision, or whether they reflect information received by the neuron about the decision formed in higher cortical areas has been a matter of debate  \cite[]{Shadlen1996, Nienborg2009, Nienborg2010}. In our model, CPs are the result of both feedforward (bottom-up) {\it and} feedback (top-down) processing. A stochastic increase in firing of a sensory neuron does increase the probability of a decision consistent with the activity of this neuron (feedforward). In the absence of noise correlations caused by the decision variable  \cite[]{Ecker2010}, the strength of this effect depends inversely on the number of sensory neurons contributing to the decision  \cite[]{Shadlen1996,Haefner2013}. In our model, CPs at the beginning of the trial are primarily due to this feedforward pathway since no coherent top-down belief has formed yet, i.e. the current belief over $D$ is only weakly correlated with the final choice. While we assume an unbiased prior over the correct decision, biased expectations about the upcoming correct choice will generally increase CPs at stimulus onset -- albeit through the feedback pathway. At any point throughout the trial, the brain's belief about the correct decision reflects the accumulated sensory information presented earlier in the trial. This constitutes prior information about the likely retinal image at that point which influences the brain's belief about the content of the image and, hence, the responses of sensory neurons representing this belief. In general, as the trial progresses and the top-down belief about the correct decision becomes stronger, CPs are enhanced by an increasing top-down component and increasingly reflect the accumulation of evidence about the decision variable (Figure \ref{fig_CP}a+c). The crucial point here is that the posterior belief over $\vx$ at some time $t$ within the trial depends on both the current observation $\vI_t$, but also all previous observations $\vI_{1..t}$. Information about these previous observations is communicated to a neuron representing $\vx_t$ via the posterior belief over the correct decision, $p_t(D)$ at time $t$ (Figure \ref{fig_CP}e, i.e. $p_t(D)\equiv p(D|\vI_1,..,\vI_t)$).

Second, CPs in our model are largest for those neurons whose preferred orientation is closest to the task-relevant orientations (Figure \ref{fig_CP}b) -- a relationship in agreement with empirical data  \cite[]{Cohen2009, Bosking2011}. Since neurons whose preferred orientation is aligned with the task axes are also the most informative about the correct decision, the relationship in Figure \ref{fig_CP}b implies a correlation between neurometric thresholds and CPs, consistent with empirical findings  \cite[]{Nienborg2012}. As for noise correlations described above, our model predictions concern the qualitative shape of the CP dependence on time and preferred orientation, not their magnitude. Regardless of the particular parameter values in our model, CPs increase over time and are largest for neurons most modulated by the task-relevant stimulus dimension. This is in agreement with empirical evidence not just for coarse but also fine discrimination tasks \cite[]{Purushothaman2005} where neurons with the steepest tuning curve slope have the highest CPs.
Third, as for the amplitude of the noise correlations, the magnitude of CPs is related to the degree to which the brain has learnt the task model. This predicts that the CP for task-relevant neurons should increase with learning, as has been observed empirically  \cite[]{Law2008}.

\subsection*{Correlations between stimulus and behavior}

\begin{figure}\begin{center}
  \includegraphics[width=12cm]{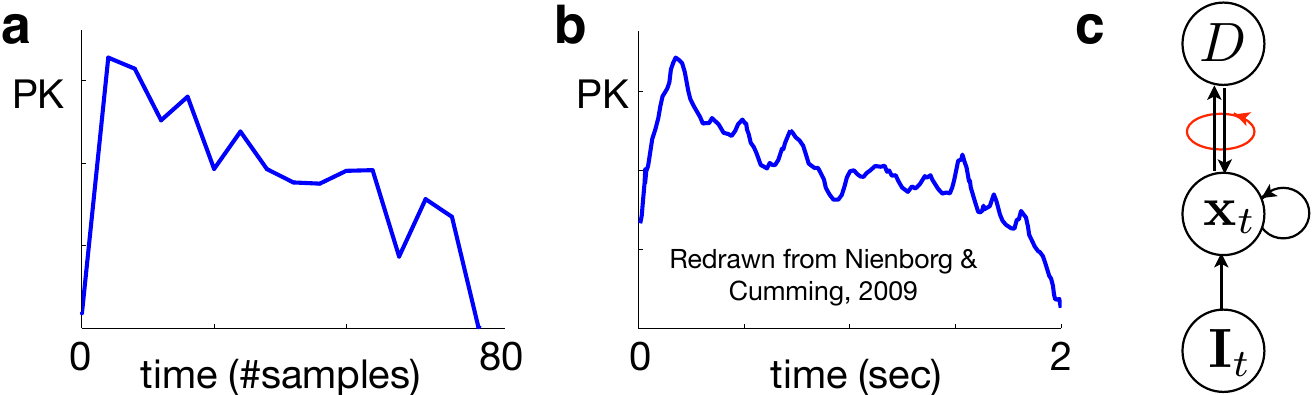}
\caption{Time dependency of psychophysical kernel (PK). 
{\bf a:} Model prediction. Evidence early in the trial is weighted more strongly than evidence presented late in the trial.
{\bf b:} Empirical PK in a disparity discrimination task  \cite[]{Nienborg2009}.
{\bf c:} 
Information flow in our model: feedforward from the retinal image $\vec{I}_{t}$ to the sensory representation $\vx_t$, and from the sensory representation to the decision $D$. Feedback from decision to sensory representation, and lateral within the sensory representation (for equations see SI). Feedforward and feedback between sensory representation and decision area form a self-reinforcing loop entailing a decreasing PK.
}
\label{fig_PK}
\end{center}\end{figure}

The strength of the correlation between stimulus and behavior in 2AFC tasks is typically measured by the psychophysical kernel (PK) \cite[]{Neri1999, Ahumada2002, Nienborg2009}. The PK quantifies how strongly the evidence in the stimulus is weighted in the decision-making as a function of the time at which the evidence is presented during the trial. Our model predicts that the weighting decreases over time (Figure \ref{fig_PK}a) so that evidence presented early in the trial has a larger influence on the final decision than evidence presented near the end. This order effect is not inherent to probabilistic inference -- exact inference leads to a constant PK as would be expected from the optimal solution -- but is a consequence of performing approximate online inference (sequential sampling in our model, also see SI). In our model, the reason for this decrease is a self-reinforcing feedback loop involving sensory and decision neurons (Figure \ref{fig_PK}c). In the absence of any stimulus information, when the responses of the sensory neurons are entirely determined by their prior, small deviations from an exactly even (50-50) belief over $D$ will bias the brain's belief about $\vx$ by top-down belief propagation. This bias in the sensory representation in turn reinforces the initial deviation in the brain's belief over $D$ in a feedforward fashion. Even if there is information in the stimulus, as long as it does not dominate the top-down prior -- as is usually the case in threshold psychophysics -- this mechanism leads to early evidence having a larger impact on the final choice than evidence presented later in the trial. Furthermore, the strength of this order effect should depend on how strong the influence of the bottom-up stimulus evidence is on the sensory representation compared to the top-down prior. 

Both predictions are in agreement with empirical data. A decreasing PK was found in the same coarse discrimination task (Figure \ref{fig_PK}b) that reported CPs to be increasing over time (Figure \ref{fig_CP}c) \cite[]{Nienborg2009}. Here, the stimulus consisted of weak disparity signals embedded in noise. A recent study in rats using a Poisson-clicks stimulus, on the other hand, found a constant PK when the task was to report whether more clicks were played to the left or the right ear \cite[]{Brunton2013}. Here, the individual clicks were highly super-threshold such that the influence of the top-down prior compared to the bottom-up likelihood was presumably negligible. Therefore, we predict that as the volume of the clicks is reduced to detection threshold level, or if the clicks are embedded in distractor noise, the PK for the Poisson-clicks task should become downward-sloping, too.  We emphasize that the model was not designed to contain this self-reinforcement of beliefs but that this feature emerged as a side-effect of performing probabilistic inference in an online fashion using a neural sampling based implementation in a hierarchical model. Its mathematical form is determined by the structure of the task and it shares the same parameters that determine magnitude and time course of noise correlations and CPs.

\section*{Discussion}

In this paper we have demonstrated how the structure of a psychophysical task can be used to make empirically testable predictions for neural responses from the old idea of visual perception as probabilistic inference even though the brain's internal model for general vision is unknown. We have applied these ideas to a simple 2AFC task and found that our model naturally accounts for several findings that are hard to explain in the traditional feedforward framework and makes new predictions for neurophysiology and psychophysics. The crucial difference to previous models of this task -- probabilistic  \cite[]{Beck2008} or not  \cite[]{Gold2007} -- was our assumption that the brain performs inference not only about the task-relevant variables but also the variables represented by sensory neurons. Strictly speaking, to perform a given task, one only has to compute the posterior over the task-relevant variable(s). However, in general, these variables could be (and in real-life usually are) many, including low-level as well as high-level features represented at different levels of the visual hierarchy, and furthermore, these variables are usually not even pre-specified. Decision-making in such a context requires inference over all latent/unobserved variables, not just a subset pre-defined by the experimenter. This assumption implies that the responses of sensory neurons represent posterior beliefs, i.e. not just certain stimulus features but also incorporating prior information from the rest of the brain \cite[]{Lee2003}. This provides a functional constraint on the top-down connectivity that is heavily under-constrained by empirical data and, therefore, often ignored (but see \cite[]{Rao1999, Wimmer2015}). Since the experimenter controls the task that determines the functional form of this top-down influence in our normative model, the validity of this normative constraint can be tested very directly. 
The first new and specific prediction of our model, about the structure of the full correlation matrix in primary visual cortex during a coarse orientation discrimination task (Figure \ref{fig_Corr}a), has since been confirmed by independent preliminary evidence (Bondy \& Cumming, SfN 2013 Conference Abstract).

\begin{figure}\begin{center}
  \includegraphics[width=14cm]{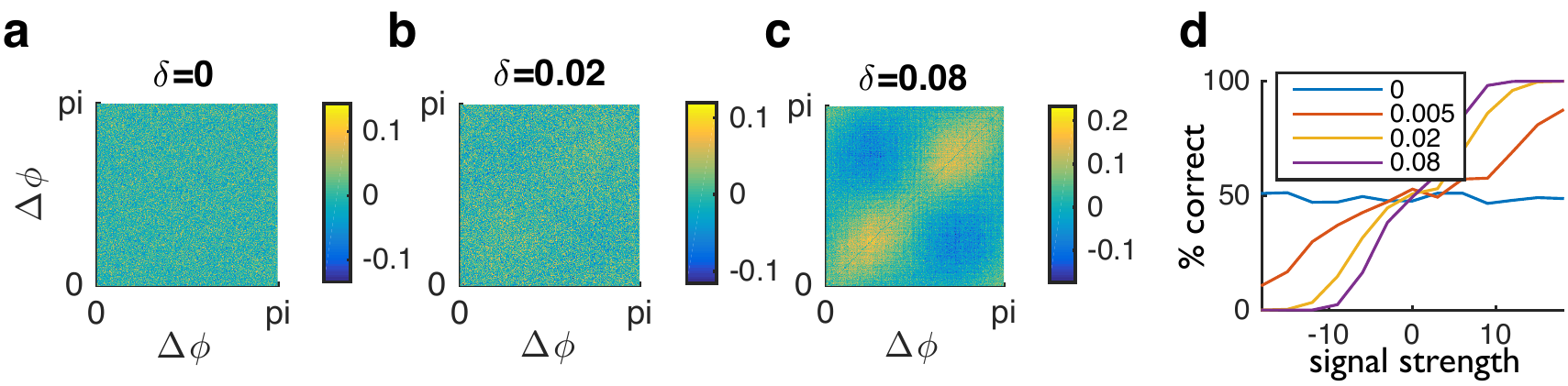}
\caption{Relationship between amplitude of noise correlation structure and psychophysical performance: stronger correlations reflect higher performance.
{\bf a-c:} Correlation structure for three models that differ in the strength of the bottom-up and top-down connections (quantified by parameter $\delta$ (see Methods), indicated on top of each panel). As $\delta$ increases from 0 (no learning), the correlation structure reflecting the 2AFC task emerges. Note the different color scales in the three panels.
{\bf d:} Psychometric curves for 4 models. At $\delta=0$ performance is at chance, and as $\delta$ increases, the curves become steeper, i.e. performance increases. The legend indicates the value of $\delta$ for each of the simulated models.}
\label{fig_Corr_Performance}
\end{center}\end{figure}

An important feature of our model prediction for the noise correlation matrix between sensory neurons is that its principle structure (2 maxima and 2 minima and their location) is the result of a normative approach and was not fit to the data we aimed to explain. We assumed that the brain performs the task using probabilistic inference. We did {\it not} assume that the brain does so optimally and our approach makes explicit the way in which the brain deviates from optimality. Suboptimality in our model is the consequence of three features. First, sampling-based approximations of probabilities converge to the exact solution only in the limit of infinitely many samples. Since the brain can only generate a finite number of samples, its solution to any problem will deviate from the optimal solution. Second, in conjunction with an online processing of the evidence, we show this to lead to an overweighting of early evidence -- another deviation from optimality. And third, the internal model that the brain has learnt about the task will generally deviate from the true external, experimenter-defined one. These deviations from optimality are made explicit by our model and, in fact, gave rise to its free parameters: the number of samples generated per trial, the strength of the brain's belief that each rewarded decision is preceded by a grating of a particular orientation in the stimulus, and the strength of the belief that a particular grating is characterized by a particular response of sensory neurons. At one extreme of the parameter range, the brain has not learnt the task at all and the task, therefore, has no impact on sensory responses: task-dependent noise correlations, CPs and PK would then be zero. The reason for this is that in the absence of a task, our model only consists of a fixed sparse coding model for V1. However, we remain agnostic about the non-task-related correlations in sensory neurons and, strictly speaking, our model only makes predictions for the difference between before and after task-learning. At the other extreme of the parameter range, the internal model of the brain becomes identical to the experimenter's model at which point the psychophysical performance of the model becomes optimal given the constraints of the V1 part of the model (which we assume to be fixed) and the online processing constraint. A direct prediction from our modeling is, therefore, that task-induced CPs and correlations should increase during task-learning (Figure \ref{fig_Corr_Performance}). We emphasize that our qualitative predictions hold across the entire model parameter regime: independent of the precise values for the parameters, CPs increase over time, PKs decrease, and the noise correlation structure has two maxima and two minima whose locations are defined by the task-relevant stimuli. 
At first sight, our prediction that the amplitude of the predicted correlation structure should increase with perceptual learning appears to be at odds with the finding that perceptual learning decreases noise correlations in visual and vestibular tasks in area MSTd  \cite[]{Gu2011}. However, that reduction was only seen for the {\it average} correlations (for which our model makes no prediction) and not for the slope of the relationship between noise correlations and signal correlations which is only loosely related to the amplitude of the task-dependent correlation matrix.

The role of correlations between sensory neurons in the probabilistic inference framework is very different from that in the traditional feedforward information processing view. While noise correlations typically limit the information about the external stimulus that can be represented by populations of sensory neurons  \cite[]{Zohary1994, Averbeck2006a, Moreno-Bote2014}, or at least complicate the read-out  \cite[]{Shamir2006, Ecker2011}, in the probabilistic inference model correlations reflect prior information about the structure of the outside world  \cite[]{Berkes2011}. This information is communicated to the relevant sensory neurons and, hence, modifies their responses. This belief propagation is best seen in the increasing time course of CPs which mirrors the formation of the brain's cognitive belief about the correct decision (Figure \ref{fig_CP}a+c). 

While attention and belief-propagation have been argued to employ the same biological mechanism  \cite[]{Krug2004}, they make different neurophysiological predictions in a 2AFC task. In order to increase psychophysical performance in the traditional feedforward framework, attention needs to increase the responses of neurons supporting both choices {\it on the same trial}. Otherwise -- enhancing the responses of neurons supporting one choice only -- will, in the absence of any external attention cue that is valid more often than not (a cue that is not present in the experiments we model), simply lead to biased decision-making decreasing psychophysical performance. Belief-propagation, on the other hand, increases the responses of only those neurons that support the choice believed to be more likely by the rest of the brain.
Only the latter mechanism, but not the ``performance-improving kind of attention'', leads to CPs. Furthermore, a gain change that varies from trial to trial but acts equally on neurons supporting either choice (as required in order to improve performance), implies equal noise correlations independent of which choice the neurons support -- in contrast to empirical findings  \cite[]{Cohen2008}. (Also, the implied correlation matrix would have peaks where our prediction shows troughs.) Finally, since an alternating of attention would reduce performance and lead to an inverse relationship between correlation strength and performance \cite[]{Ecker2015} (in contrast to Figure \ref{fig_Corr_Performance}),
we argue that belief propagation is the more parsimonious explanation for the observed task-dependent correlations.

The observation that the time courses of CPs and PKs differ during the stimulus presentation has been a challenge to feedforward models of sensory processing and has been taken as evidence for a feedback component in CPs  \cite[]{Nienborg2009}. These findings are fully consistent with our model where CPs contain both a feedforward and a feedback component. The existence of a feedforward component can most easily be seen by the fact that CPs are larger than 0.5 at stimulus onset when the model's belief about the correct decision is still $50-50$. The magnitude of this bottom-up component follows the same decreasing time course as the PK (Figure \ref{fig_PK}a)  \cite[]{Nienborg2009} with the difference accounted for by top-down belief propagation. Furthermore, we have demonstrated that the very same feedback signal that causes the increasing CPs also causes a decreasing PK
due to a positive feedback loop between decision-making neurons and sensory neurons. We stress that we did not hand-craft the model structure in order to fit these observations, but that this feedback signal is a direct result of performing probabilistic inference in this task.

There exist alternative explanations for a decreasing PK suggesting that the brain inappropriately uses strategies optimal for reaction-time tasks ('integration-to-bound'  \cite[]{Gold2007}) or hypothesizing costs intrinsic to accumulating evidence  \cite[]{Drugowitsch2012}. However, our explanation demonstrates that even if a neural decision circuit \emph{itself} equally integrates the signals it receives from sensory neurons over time as has been suggested recently  \cite[]{Brunton2013}, its top-down influence on the very same sensory neurons 
can lead to early evidence being weighted more strongly than evidence presented at the end of the trial if that evidence is weak. Future work on extending our model to a reaction time paradigm and comparing its predictions with reaction-time data will be able to assess the relative importance and explanatory power of these different hypotheses.

Our study complements a growing body of literature suggesting that sampling-based probabilistic inference may underly higher perceptual and cognitive processes  \cite[]{Griffiths2012, Vul2014} and it suggests that probabilities might also be represented by samples in lower sensory processing (given the compatibility with the data considered here). It is particularly intriguing that basic visual perception appears to be susceptible to the same confirmation bias that is ubiquitous in the context of higher cognitive reasoning  \cite[]{Nickerson1998}. The same two factors that contribute to a decreasing PK in our model -- a sequential sampling approximation and evidence accumulation on the basis of inferred beliefs instead of directly observed information -- appear to be present in other contexts as well. This suggests sequential sampling-based evidence accumulation as a new candidate mechanism underlying confirmation bias, different from previous explanations  \cite[]{Lieder2012}. More generally, our framework bridges cognitive science and systems neuroscience by constructing a rational process model  \cite[]{Griffiths2012} combining the generative model for the psychophysical task with knowledge about the biological architecture and the neural sampling hypothesis.

Our work differs from a previous probabilistic model  \cite[]{Beck2008} of the 2AFC task in two important aspects. First, we assume that cortical sensory neurons represent unobserved or inferred variables. This aspect of our model requires feedback connections in order to include extra-sensory information in the computations performed by sensory neurons, connections that are missing from \cite[]{Beck2008}. Without this feature of our model there would be no task-dependent noise correlations, nor would there be CPs that increase while the PK decreases. Second, our model differs from that of Beck et al. in that the representation of probabilities is sampling-based  \cite[]{Fiser2010, Hoyer2003} rather than based on a 'probabilistic population code' (PPC)  \cite[]{Ma2006}. Preliminary evidence suggests, however, that our results generalize to a PPC-based representation (A. Pouget -- private communication.)

It has previously been suggested that priors may also been encoded in feedforward weights \cite[]{Ganguli2011,Wei2012}. While this is likely for long-term priors reflecting permanent statistics of the natural environment, it appears impossible for the task-dependent context information considered in our study that can vary on a trial-by-trial basis \cite[]{Cohen2008}; or indeed the influence of the stimulus information at the beginning of a trial on later sensory beliefs within the same trial (Figure \ref{fig_CP}e).

A recent paper \cite[]{Wimmer2015} has presented a mechanistic model that reproduces an increasing CP and a decreasing PK. There are similarities between the computational motifs of mutual inhibition and a positive feedback loop between sensory and decision neurons in their model and in ours. However, unlike our model theirs is constructed in a bottom-up/data-driven fashion, and therefore cannot address the question of 'why' the brain is structured the way it is -- especially since mutual inhibition ('explaining away' in our model) and feedback connections ('belief propagation') decrease the psychophysical performance of their model. Furthermore, their model is designed specifically for the 2AFC task under consideration, without relating it to the general probabilistic inference problems that the brain has to solve during natural vision.

A major open question in systems neuroscience concerns the origin and role of correlated neuronal variability. It has long been known that neuronal responses are correlated, and the development of population recording techniques is finally making detailed measurements of these correlations feasible, in particular their magnitude, shape and dependency on both stimulus and cognitive variables. 
In the framework of probabilistic inference, the generative model for the stimulus in the task acts as a prior on the variables represented by the sensory responses. 
Since the generative model for the task is entirely under the control of the experimenter this allows for designing experiments with very direct predictions. We therefore hope our results can inspire future experiments, and provide a framework for analyzing multi-neuron recordings in the awake brain and for linking these recordings to otherwise hard-to-measure cognitive factors like beliefs and task strategies.

\section*{Experimental Procedures}

\subsection*{Generative model learnt by the brain}

We assume that the brain has learnt the two crucial features of the coarse orientation discrimination task that we are modeling: that there are exactly two possible decisions, $D=1$ and $D=2$, and that they are related to the orientation of the visual stimulus presented on the screen. We further assume that the subject has learnt that both choices are equally likely a priori, $p_{D}(D=1)=p_{D}(D=2)=0.5$. From correlating correct choices with the preceding stimulus, the brain has to learn the task-relevant orientations $\psi_1$ and $\psi_2$. Inevitably, there will be some level of uncertainty (tolerance) about the precise orientations indicating choice $1$ and choice $2$. We model this uncertainty by a circular Gaussian (von Mises) function around the correct orientation:
$p(g_i=1|D=k)	\;\propto\;	\exp[\kappa\cos 2(\phi_i^{(\vg)}-\psi_k)]$
where $g_i$ is one of $\nG$ binary variables indicating the presence of a grating with orientation $\phi_i^{(\vg)}$ in the stimulus. The normalization 
is chosen to be $\sum_i^\nG p(g_i=1|D=k)=1$, such 
that on average one of the grating variables is 'on'. $\kappa$ is a free parameter in our model that can range from $0$ to $\infty$ where $\kappa=0$ represents no knowledge about which orientations are task-relevant and $\infty$ represents the limit of perfect learning, where the internal model is identical to the generative model used by the experimenter. We assume that the task is performed on the basis of an early sensory representation ($\vx$, here primary visual cortex) such that each orientation variable $g_i$ corresponds to a particular expected value for $\vx$. Again, we make a canonical circular Gaussian assumption about the dependency between a grating variable $g_k$ representing orientation $\phi_k^{(\vg)}$ and a Gabor-shaped feature $x_i$ with orientation $\phi_i^{(\vx)}$ in the image:
\begin{align}
\E[x_i|\vg]	=	1+\delta\sum_{k=1}^\nG g_k\exp\left[\lambda\cos 2\left(\phi_i^{(\vx)}-\phi_k^{(\vg)}\right)\right].
\label{eq_Ex}
\end{align}
Here, $0<\lambda$ and $0<\delta$ are free parameters (discussed below). Since the $x_i$ are meant to model neurons in V1, we model their relationship to the image on the retina by a standard sparse-coding model  \cite[]{Olshausen1997, Hoyer2003}:
\begin{align}
p(x_i|\vg)	&=	\frac{1}{\tau_i}\exp(-x_i/\tau_i)\,H(x_i)\\
p(\vI|\vx)	&=	\mathcal{N}\left(\vI:\frac{1}{\nx}\sum_{i=1}^\nx\vec{PF}_i x_i,1\right)
\end{align}
with $\tau_i\equiv\E[x_i|\vg]$ as defined in equation (\ref{eq_Ex}) and $H$ being the Heaviside function restricting $\vx$ to positive values. (The $x_i$ can be interpreted as firing rates here with no specific assumption about the spike production mechanism.) The matrices $\vec{PF}_i$ contain the projective fields for each $x_i$ and the actual image on the screen is a noisy version of their linear superposition weighted by $\vx$. Since the only stimulus dimension relevant for our results is orientation, we simply assume a bank of Gabor-shaped filters (aspect ratio 2:1) which only differ in their orientation. For $\delta=0$, the model for the sensory responses becomes independent of the task and the prior over $\vx$ is sparse and independent. As $\delta$ increases, the expected intensity of the pattern associated with the presence of a grating in the stimulus increases. The generative model described here determines the sampling equations which are provided in the supplementary materials (\ref{S_Eqs})

The generative model described above will be normative, i.e. correct, only for stimuli that have actually been generated from it. Despite its deviations from the models actually used by the particular experiments that we compared our data to  \cite[]{Cohen2008,Nienborg2009,Brunton2013} it captures the two crucial features of a 2AFC orientation discrimination task: 1) the expectation of an oriented stimulus of some orientation, and 2) the expectation that only one of two possible orientations (or ranges of orientations) will be shown. $\kappa$ and $\lambda$ together with the Gabor-shaped filters determine the orientation bandwidth of the expected oriented stimulus and $\delta$ can be interpreted as the expected saliency or contrast of the signal in the noise. 

In tasks where the stimulus is dynamically changing within a trial, evidence about the correct decision has to be accumulated over time. In this case, the brain will learn that while there is only one correct choice per trial, the image and latent variables other than $D$ can change from stimulus frame to stimulus frame. As a result the posterior over $D$ evolves over time starting with the initial prior $p_0(D)$, with $p_t(D)\sim p_{t-1}(D)p(\vI_t|D)$ for independent $\vI_{t-1}$ and $\vI_t$  \cite[]{Beck2008}. Such an accumulation of evidence is compatible with existing neurophysiological observations in putative decision-making regions of the brain  \cite[]{Gold2007} providing a graded representation of the instantaneous belief over the correct decision $D$. In our hierarchical model the posterior over $D$ is not updated based on the observed variable $\vI$ directly but based on the samples drawn from the posterior over $\vg$, $p(\vg|\vI_t)$. We denote the number of samples generated on the time scale on which the $\vI_t$ are assumed to be statistically independent by $n_s$. The belief over $D$ is accordingly updated in an online fashion based on individual samples: $p_t^{(k)}(D)\sim p_t^{(k-1)}(D)p(\vg_t^{(k)}|D)^{1/n_s}$ with $k=1..n_s$ and $p_t^{(1)}(D)\equiv p_{t-1}^{(n_s)}(D)$, and where $\vg_t^{(k)}$ are the corresponding samples drawn from $p(\vg_t|\vI_t,D)$. $n_s$ represents the time scale on which sensory evidence in the external world is assumed to be independent by the brain and constitutes another free parameter in our model. Experimentally, it can be constrained by the time course of the CP: the smaller $n_s$ the faster the temporal increase in CP. 

In our simulations we explicitly model the $\vx$ and $\vg$ levels. We do not implement a spike-based representation of the belief over $D$ but simply assume that an area exists in the brain (e.g. LIP \cite[]{Gold2007, Beck2008}) which accumulates the sensory evidence and represents the brain's current belief over the correct decision. This belief over $D$ acts as a top-down prior for the sensory representation, as required by full probabilistic inference over all latent variables, i.e. including $\vx$ and $\vg$, and not just the decision variable $D$. We note that two features responsible for the decreasing PK are easily overcome in typical machine learning applications. First, if the updating of the posterior over $D$ is not performed after every sensory sample, but in 'batches', e.g. after every 10 samples, then evidence is weighted more equally over time. We do not think this is biologically plausible since it would require the brain to cache incoming samples and effectively ignore them during the caching time, before updating the posterior -- without explicit knowledge when would be the right time for an update of its belief over $D$. Instead, we believe that it is more plausible that evidence is continuously integrated. Second, generating more than one chain of samples in parallel (e.g. particle filtering \cite[]{Doucet2001}) would also prevent the PK from decreasing and has previously been suggested \cite[]{Lee2003}. Investigating the source of the decreasing PK in more detail as suggested in the Discussion section will therefore give further insights into what kind of sampling scheme the brain is implementing.

\subsection*{Numerical details}

We performed Gibbs sampling in the generative model (see \ref{S_Eqs}). Trials start after the burn-in period and last 80 complete samples, i.e. 80 samples are generated from each variable in the model. For the simulations underlying Figures \ref{fig_Corr}-\ref{fig_PK} we used: $\kappa=1$, $\lambda=3$, $\delta=0.016$, $\nx=1024$ and $\nG=256$. The only exception is the 3AFC Figure \ref{fig_Corr_2}a where we used $\kappa=3$, i.e. a narrower range of orientations related to each choice. We further assumed 80 samples per trial (i.e. 80 updates of each latent variable in the model), and $n_s=20$. Since these are free parameters in our model, we concentrate on qualitative rather than quantitative predictions. However, when modeling a particular experiment, they can be constrained by fitting them to observations, in particular CPs, noise correlations and classification images.

The noise correlations and CP time course were computed using a grey screen as stimulus in order to eliminate stimulus-induced influences and to avoid having to correct CPs for them  \cite[]{Nienborg2009}. The PK time course was computed using a dynamic stimulus in which randomly switching gratings were embedded in Gaussian noise. 16 independent stimulus frames were presented per trial.

\subsection*{Choice probability (CP) and psychophysical kernel (PK)}

The CP of a particular neuron with respect to choice 1 can be defined as the probability that a random sample from the neuron's response distribution preceding choice 1 is larger than a sample from the same neuron's response distribution preceding choice 2 \cite[]{Britten1996}.

The PK can be defined as the amplitude of the classification image as a function of time \cite[]{Nienborg2009}. The classification image is the difference between the mean stimulus preceding choice 1 and the mean stimulus preceding choice 2 \cite[]{Neri1999, Ahumada2002}. In order to compute the time course of the PK, one computes the classification image (essentially the choice-triggered average of the stimulus) as a function of time \emph{within} a trial.

\subsubsection*{Acknowledgments}
We sincerely thank the many colleagues with whom we have discussed this work and who have provided us with valuable advice on content and presentation. RMH acknowledges financial support by the Swartz Foundation.

\small 
\bibliography{Paper_Sampling_Decision}
\bibliographystyle{cell} 

\newpage

\setcounter{page}{1}
\setcounter{figure}{0}
\setcounter{section}{1}
\setcounter{equation}{0}
\renewcommand{\thefigure}{S\arabic{figure}}
\renewcommand{\thepage}{S\arabic{page}}
\renewcommand{\thesection}{S\arabic{section}}
\renewcommand{\theequation}{S\arabic{equation}}

\section*{Supplementary Information}

\subsection{Sampling equations}\label{S_Eqs}

To simplify notation, we write each image $\vI$ as a column vector $\vy$, and combine the set of rescaled projective fields $\{\frac{1}{\nx}\vec{PF}_i\}$ into a matrix $\vec{G}$. The generative model equations become:
\begin{align}
D		&\sim	\text{Categorical}(\pD)\quad\text{where $D$ is 1 or 2}\\
g_k|D	&\sim	\text{Bernoulli}\left\{\frac{1}{\nG I_0(\kappa)}\exp\left[\kappa\cos 2(\phi^{(\vg)}_k-\psi_D)\right]\right\}\\
p(x_i|\vg)	&=	\frac{1}{\tau_i}\exp(-x_i/\tau_i)\,H(x_i)\\
		&\quad\text{with } \tau_i=1+\delta\sum_{k=1}^\nG g_k
			\exp\left[\lambda\cos 2(\phi_i^{(\vx)}-\phi_k^{(\vg)})\right]\label{eq_tau_S}\\
p(\vy|\vx)	&=	\mathcal{N}\left(\vy:\vec{Gx},\sigma_\vy^2\right)
\end{align}
where $I_0$ is the modified Bessel function of zeroth order, and $H(x)$ is the Heaviside function.

From this follow the conditional probabilities for Gibbs sampling:
\begin{align}
p(D|\vg)	&\Propto	p(\vg|D)\pD(D)\\
		&\Propto	p_D \prod\limits_{j=1}^\nG
			\begin{cases}
			\frac{1}{\nG I_0(\kappa)}\exp\left[\kappa\cos 2\left(\phig_j-\psi_D\right)\right]	& \text{for } g_j=1\\
			1-\frac{1}{\nG I_0(\kappa)}\exp\left[\kappa\cos 2\left(\phig_j-\psi_D\right)\right]	& \text{for } g_j=0
			\end{cases}\\
p(g_k|D,g_{\neg k},\vx)	&\Propto	p(\vx|\vec{g})p(g_k|D)\\
			&=\begin{cases}
			\frac{1}{\nG I_0(\kappa)}\exp\left[\kappa\cos 2\left(\phig_j-\psi_D\right)\right]	& \text{for } g_j=1\\
			1-\frac{1}{\nG I_0(\kappa)}\exp\left[\kappa\cos 2\left(\phig_j-\psi_D\right)\right]	& \text{for } g_j=0
			\end{cases}\\
			&\quad\times
			\prod\limits_{j=1}^\nx \frac{1}{\tau_j}\exp[-x_j/\tau_j]
			\quad\text{with $\tau_j$ defined as in equation (\ref{eq_tau_S}})\\
p(x_k|\vg,x_{\neg k},\vy)	&\Propto	p(\vy|\vx)p(x_k|\vg)\\
			&\Propto	\exp\left[-\frac{(\vy-\vec{Gx})^\top(\vy-\vec{Gx})}{2\sigma^2_\vy}\right]
					\exp\left[-\frac{x_k}{\tau_k}\right]H(x_k)\\
			&\Propto	\exp\left[-\frac{\vy^\top\vy-2\vy^\top\vec{Gx}-\vx^\top\vec{Rx}}{2\sigma^2_\vy}-\frac{x_k}{\tau_k}\right]H(x_k)\\
			&{}	\text{where we have defined: $\vec{R=-G^\top G}$}
\end{align}
This means $x_k$ is drawn from a cut-Gaussian distribution  \cite[]{Hoyer2003}
with mean and variance:
\footnote{Matching of coefficients of $\frac{-(x-\mu)^2}{2\sigma^2}$ with $Ax^2+Bx+C$ requires that $\sigma^2=-\frac{1}{2A}$ and $\mu=\sigma^2B=-\frac{B}{2A}$. Here, $A=R_{kk}(2\sigma^2_y)^{-1}$ and 
$B=(2\vy^\top\vec{G}_{:k}+\sum_{i\ne k}R_{ik}x_i-2\sigma^2_y\tau_k^{-1})(2\sigma^2_y)^{-1}$
}
\footnote{Note: $\vec{x^\top Rx}=\sum_i R_{ii}x_i^2+\sum_{i\ne j}R_{ij}x_i x_j$}
\begin{align}
\mu	&=	\frac{\vy^\top \vec{G}_{:k} + \vec{R}_{k,\neg k}\vx_{\neg k} -\sigma_y^2\tau^{-1}}{-R_{kk}}
	\equiv	\frac{\sum_i y_i G_{ik} + \sum_{i\ne k}R_{ik}x_i -\sigma_y^2\tau^{-1}}{-R_{kk}}\\
\sigma^2	&=	-\frac{\sigma^2_y}{R_{kk}}
\end{align}

We used 1x1 (arbitrary units) image patches of size 32x32 pixel. The PFs consisted of identical 2d-Gabor functions that only differed in their orientation (equally spaced). Their amplitude was 1, spatial frequency 2, and phase 0. The Gaussian envelope had an standard deviation of 0.1 in the phase-modulated direction, and 0.2 perpendicular to it. Each PF was rescaled such that its 2-norm was 1.

\subsection{Correlations between variables in the internal model imply correlated neural responses}\label{S_Corr}

\begin{figure}\begin{center}
  \includegraphics[width=10cm]{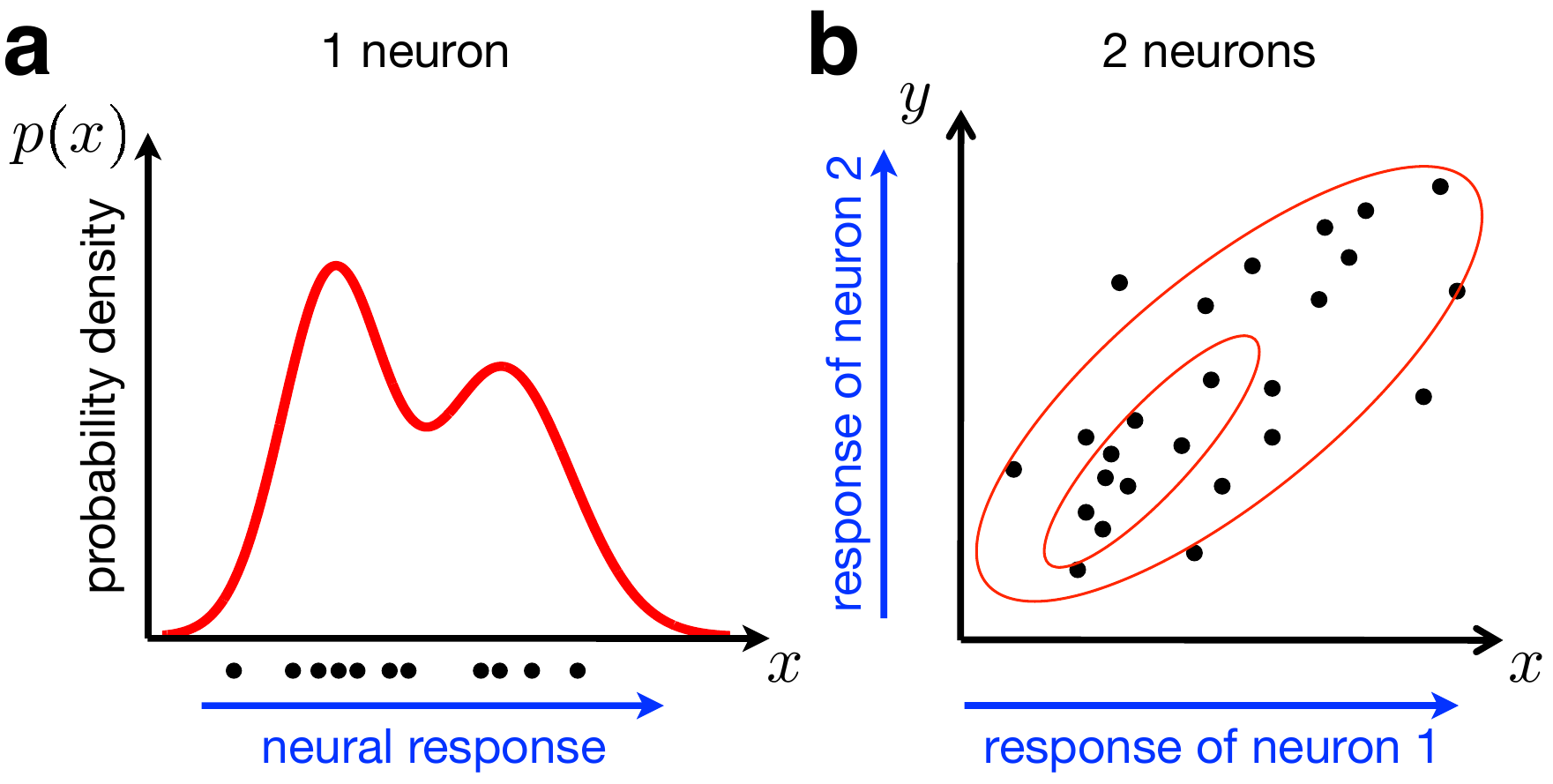}
\caption{Neural sampling based representation. 
{\bf a:} Representation of the posterior over a scalar variable $x$ by the responses of a single neuron. Response variability is related uncertainty in knowledge about $x$. 
{\bf b:} If the posterior distribution $p(x,y)$ is correlated, this implies correlations between the neural responses by the two neurons representing $x$ and $y$, respectively.
}
\label{fig_S_Sampling_Corr}
\end{center}\end{figure}

Figure \ref{fig_S_Sampling_Corr} illustrates that dependencies between two variables in the joint posterior will lead to correlated variability in their responses.

\subsection{Notes on model mismatch and sampling process}\label{S_Mismatch}

In the main text we discussed the sampling process and the reasons for, and effects of, the mismatch between the internal model that the brain performs inference in and the external model defined by the experimenter. The model shown in Figure \ref{fig_GenModel}b is easy enough to develop an analytical and conceptual understanding. Since $D$ and $g_i$ are binary variables, we can fully specify this model by:
\begin{align}
p(D=1)=p(D=2)	&=	0.5\text{  assuming no bias}\\
p(g_1=1|D=1)=p(g_1=0|D=2)	&=	a\text{  with }1/2\le a\le 1\\
p(g_2=1|D=2)=p(g_2=0|D=1)	&=	a\\
p(\vI|g_1,g_2)	&=	\mathcal{N}(\vI:g_1\vec{PF}_1+g_2\vec{PF}_2,\vec{1}).
\end{align}
From these equations it follows
\begin{align} 
p(g_1=0|D=1)=p(g_1=1|D=2)	&=	1-a\text{  and}\\
p(g_2=0|D=2)=p(g_2=1|D=1)	&=	1-a.
\end{align}
Here, $a$ parametrizes the mismatch between this model (representing the internal model) and the external, experimenter-defined model. For $a=1$, both models are identical while values smaller than 1 represent an increasing mismatch between the two. Hence the posterior, $p(D,g_1,g_2|\vI)$ computed by any inference method will be correct with respect to the actual task, and behavioral performance based on it, as measured by \% correct, for instance, will be maximal. On the other hand, $a=0.5$ represents maximal mismatch within the context of this model, and performance will be minimal. In fact, it is easy to see that $p(D|\vI)=0.5$ regardless of the input $\vI$ in the case of $a=0.5$ meaning the model performance will be at chance level. Furthermore, as described in the main text, for $a=1$, the posterior will only have mass at $(g_1=1,g_2=0)$ and $(g_1=0,g_2=1)$ implying a perfect anti correlation between the samples for $g_1$ and $g_2$. Again, at the other extreme for $a=0.5$, it is easy to see for an ambiguous stimulus $\vI$ (e.g. pure noise as used when computing CPs and noise correlations, or equal agreement with $\vec{PF}_1$ and $\vec{PF}_2$) that $p(g_1,g_2)$ is the same for all values for $g_1$ and $g_2$. This means that the samples for $g_1$ and $g_2$ will be uncorrelated. And in both cases -- performance and sample correlations -- intermediate values $1/2<a<1$ will interpolate between the respective extremes.

\subsection{Notes on a sampling-based evidence accumulation}\label{S_PK}

Consider a coarse orientation discrimination task framed in terms of the generative model in Figure \ref{fig_CP}e. We want to infer the state of a binary orientation variable $D$ given a series of images i.i.d. drawn from a distribution $p(I|D)$. The task is to infer $D$ given a sequence of images:
\begin{align}
p(D|I_1,\ldots,I_n) &\Propto p(I_1,\ldots,I_n|D)p(D)\\
&= p(D)\prod\limits_{i=1}^np(I_i|D)
\end{align}
Let $p_t(D) \equiv p(DI_1,\ldots,I_n)$, then the following update rule is still exact:
\begin{align}
p_{t}(D)	&\Propto p_{t-1}(D)p(I_t|D)\\
		&= p_{t-1}(D)\int p(I_t,x|D)\rd x\\
		&= p_{t-1}(D)\int p(I_t|D,x)p(x|D)\rd x\\
		&= p_{t-1}(D)\int p(I_t|x)p(x|D)\rd x \label{eq_update_integral}
\end{align}
In a sampling-based representation, this integral can be computed approximately as
\begin{align}
\int p(I_t|x)p(x|D)\rd x	&\approx	\frac{1}{n_{s}}\sum\limits_{k=1}^\ns\frac{p(I_{t}|x^{(k)})p(x^{(k)}|D)}{p_{s}(x^{(k)})}
\end{align}
where the samples $x^{(k)}$ are drawn from some sampling distribution function $p_{s}(x)$. Assuming that the brain only has access to the samples that it has generated from the posterior over $x$, $p_s(x)=p(I_{t}|x)p(x|D^{(k)})$ yields
\begin{align}
\int p(I_t|x)p(x|D)\rd x	&\approx	\frac{1}{n_{s}}\sum\limits_{k=1}^\ns\frac{p(x^{(k)}|D)}{p(x^{(k)}|D^{(k)})}.
\label{eq-integral-sample}
\end{align}
Further assuming that the brain updates its belief continuously, i.e. after each sample, leads to
\begin{align}
p_{t}^{(k)}(D)	&\appropto	p_{t}^{(k-1)}(D)p(x^{(k)}_{s}|D)^{1/\ns}
\end{align}
with $p_t^{(1)}(D)\equiv p_{t-1}^{(n_s)}(D)$ as defined in the Methods. Note that $p(x^{(k)}|D^{(k)})$ gets absorbed by the proportionality constant when updating $p_{t}(D)$ after every single sample. This formulation also makes explicit that the bias is due to both the fact that the samples from $x$ are drawn from the posterior, and that the belief over $D$ gets updated after every sample. Normally, the  ``importance weight''  $1/p(x^{(k)}|D^{(k)})$ in equation (\ref{eq-integral-sample}) compensates for the fact that the samples $x^{(k)}$ do not just represent the observed information $I_t$ but are biased by the top-down prior belief $p(x|D)$ such that the computation of the integral and hence update of $p_t(D)$ in equation (\ref{eq_update_integral}) is unbiased. However, if the update is performed after every sample, this ``correction factor'' cancels out since the update is performed based on 
$\frac{\int p(I_t|x)p(x|D=1)\rd x}{\int p(I_t|x)p(x|D=2)\rd x}$. 
As a result, the effect of the prior belief on the sensory representation given by $x^{(k)}$ is not compensated for when updating the brain's belief over $D$ leading to a ``double-counting'' of the prior belief, or confirmation bias.

\subsection{Supplementary discussion}

\paragraph{Analysis of neural variance:} In our framework the variance of a sensory response is increased by both variance in the stimulus-related inputs as well as in response to variability in internal beliefs (e.g. the decision). In the context of a 2AFC task, this means that the neural response variance of a sensory neuron should be significantly smaller when conditioned on the behavioral decision which is in fact what has been observed. The fact that there is a difference between the choice-conditioned means -- the very source of the much-documented choice probabilities \cite[]{Britten1996,Nienborg2012} -- implies that response variability is higher when the responses are not conditioned on the choice. The reason we did not focus on the individual variability but instead on the correlated variability is that the latter appears a better quantity to differentiate between our and previous feedforward models. While the observed reduction in the individual variability when conditioned on choice can be explained in a traditional feedforward framework \cite[]{Haefner2013}, this is impossible for the observed task-dependence of correlations.

\paragraph{Rational process model:} Our framework constructs a 'rational process model'  \cite[]{Griffiths2012} by starting with the generative model of the task and combining it with the sequential sampling postulated by the neural sampling hypothesis. The sequential sampling algorithm implies a mechanistic process by which the brain generates these samples and the sub-optimality of a decreasing PK is a direct result of this particular choice for a process. To provide a direct link to a biophysically plausible implementation of our model we refer to prior work  \cite[]{Buesing2011, Pecevski2011, Berkes2009}.

\paragraph{Relationship to predictive coding:} In the standard versions of predictive coding \cite[]{Rao1999, Spratling2010}, sensory neurons projecting to other areas represent prediction errors. The straightforward prediction from this -- that correctly predicted responses should be diminished -- appears incompatible with a probabilistic inference account in which responses to predicted stimuli (high prior) are enhanced (high posterior). Predictive coding therefore appears to contradict both empirical CPs and the task-dependent correlation measurements -- since both indicate an enhancement in the sensory response when compatible with the internal decision/prediction.

\end{document}